\pdfoutput=1
\documentclass[a4paper,aps,pdftex,floatfix,groupedaddress,pra,showpacs,twocolumn]{revtex4}

\usepackage{amsfonts,amsmath,amssymb}
\usepackage{graphicx}
\usepackage[latin1]{inputenc}
\usepackage{pxfonts}
\usepackage{textcomp}
\usepackage{hyperref, hypernat}
\hypersetup{colorlinks,citecolor=blue,urlcolor=blue,linkcolor=red}

\newcommand{\degree}{\ensuremath{^\circ}}%
\newcommand{\eg}{e.\,g.}%
\newcommand{\ie}{i.\,e.}%

\newlength{\figwidth}
\newlength{\figwidthsmall}
\begin{document}

\title{Laser-induced 3D alignment and orientation of quantum-state-selected molecules}

\author{Iftach Nevo$^{1}$}%
\author{Lotte Holmegaard$^1$}%
\author{Jens H. Nielsen$^2$}%
\author{Jonas L. Hansen$^3$}%
\author{Henrik Stapelfeldt$^{1,3}$}%
\email{henriks@chem.au.dk}%
\affiliation{$^1$\,Department of Chemistry, University of Aarhus, DK-8000 Aarhus C, Denmark \\
   $^2$\,Department of Physics and Astronomy, University of Aarhus, DK-8000 Aarhus C, Denmark \\
   $^3$\,Interdisciplinary Nanoscience Center (iNANO), University of Aarhus, DK-8000 Aarhus C,
   Denmark}

\author{Frank Filsinger}%
\author{Gerard Meijer}%
\author{Jochen K\"upper}%
\email{jochen@fhi-berlin.mpg.de}%
\affiliation{Fritz-Haber-Institut der Max-Planck-Gesellschaft,
   Faradayweg 4-6, 14195 Berlin, Germany}


\date{\today}

\begin{abstract}\noindent%
   A strong inhomogeneous static electric field is used to spatially
   disperse a rotationally cold supersonic beam of
   2,6-difluoroiodobenzene molecules according to their rotational
   quantum state. The molecules in the lowest lying rotational states
   are selected and used as targets for 3-dimensional alignment and
   orientation. The alignment is induced in the adiabatic regime with
   an elliptically polarized, intense laser pulse and the orientation
   is induced by the combined action of the laser pulse and a weak
   static electric field. We show that the degree of 3-dimensional
   alignment and orientation is strongly enhanced when rotationally
   state-selected molecules, rather than molecules in the original
   molecular beam, are used as targets.
\end{abstract}
\maketitle

\section{Introduction}
\label{intro}

The spatial orientation of a molecule is of crucial importance for its
interactions with other molecules, atoms, or electromagnetic
radiation. Driven by the possibility of controlling or enhancing
molecular reactivity and interactions~\cite{brooks:1976:science,
   loesch:1997:jpca, herschbach:2006:epjd}, as well as by
possibilities for new research directions in molecular science,
including imaging of molecular orbitals~\cite{itatani:2004:nature} and
photoelectron angular distribution from fixed-in-space
molecules~\cite{Kumarappan:PRL100:093006,bisgaard:Science:2009}, large
efforts have been devoted to developing methods that enable control
over the orientation or alignment of molecules. Here, alignment refers
to confinement of molecule-fixed axes along laboratory-fixed axes, and
orientation refers to the molecular dipole moments pointing in a
particular direction.

In the gas phase, the electric fields from moderately intense,
nonresonant laser pulses have proven particularly useful and versatile
for manipulating the alignment of
molecules~\cite{stapelfeldt:2003:rmp}. A laser field induces an
electric dipole moment in any molecule with a polarizability
anisotropy, whether polar or not, and the interaction between the
induced dipole and the laser field itself forces the molecule to
align. In the case of a linearly polarized laser field the most
polarizable axis is aligned along the polarization direction, which is
sufficient to ensure strong angular confinement of the figure axis of
the molecules and, therefore, complete alignment control of linear and
symmetric top molecules. This is termed 1-dimensional (1D) alignment.
For asymmetric top molecules 1D alignment does not suffice to provide
complete alignment control. Rather, all three molecular axes of
polarisability must be confined to laboratory fixed axes. Such
3-dimensional (3D) alignment has been experimentally demonstrated
through the use of an elliptically polarized laser
field~\cite{larsen:2000:prl, simon_thesis, tanji:2003:pra,
   Rouzee:pra:2008} or by combining two linearly polarized laser
pulses with orthogonal polarizations~\cite{underwood:2005:prl,
   Lee:2006:prl, Viftrup:prl:2007, Viftrup:pra:2009}.

In the case of polar molecules confinement of the molecular axes is
not sufficient to achieve full orientational control; it is also
necessary to control the direction of the permanent dipole. As
suggested by Friedrich and Herschbach a decade
ago~\cite{friedrich:1999:jcp, friedrich:1999:jpca}, and later
demonstrated experimentally~\cite{baumfalk:2001:jcp, sakai:2003:prl,
   tanji:2003:pra, Buck:Farnik:rev:2006, Holmegaard:PRL102:023001},
orientation can be added to alignment by combining the strong laser
field with a weak static electric field. Therefore, we use 1D
orientation to denote 1D alignment and, simultaneously, a preferred
direction of the permanent dipole moment. Likewise, 3D orientation
refers to 3D alignment occurring together with a preferred direction
of the permanent dipole moment.

Any application of aligned or oriented molecules will benefit
significantly, or simply require, that the degree of alignment or
orientation is strong. One efficient method to optimize the degree of
alignment or orientation is to ensure that the molecular sample
initially resides in the lowest lying rotational states. This is
achieved by lowering the rotational temperature as much as possible.
In practice, strong cooling can be obtained by supersonically
expanding a small fraction of the molecular gas of interest in an
inert carrier gas. Very strong degrees of alignment have been reported
using molecular beams with a rotational temperature around
1~K~\cite{kumarappan:2006:jcp}. Optimizing the degree of orientation
in the mixed laser and static electric field method imposes, however,
an even stronger requirement on the initial rotational state
distribution. Recently, we showed that 1D orientation, using the
mixed-field method, can be strongly enhanced by using
rotational-quantum-state selected molecules as
targets~\cite{Holmegaard:PRL102:023001,Filsinger:deflection:inprep}.
The state selection is obtained by passing a cold molecular beam
through an electrostatic deflector that spatially disperses molecules
according to their rotational quantum state. The purpose of the
present work is to demonstrate that the use of
rotational-state-selected molecules also enables significant
improvement of laser induced 3D alignment and mixed-field 3D
orientation.

The studies are carried out on 2,6-difluoroiodobenzene (DFIB)
molecules in the adiabatic limit, where the alignment laser pulse is
turned on and off slowly compared to the inherent rotational periods
of the molecules~\cite{stapelfeldt:2003:rmp}. In this regime the
degree of alignment and orientation will follow the envelope of the
10-ns-long applied laser pulse. In particular, the strongest degree of
alignment occurs at the peak of the pulse, which is the point in time
where all measurements in this work are recorded. From
quantum-chemistry calculations at the B3LYP~\cite{A.Becke:JCP.1993b,
   C.Lee:PRB.1988}/TZVPP~\cite{A.Schafer:JCP.1994} level using linear
response theory in the Turbomole 5.10 program
suite~\cite{R.Ahlrichs:CPL.1989}, the polarizability components of the
DFIB molecules are determined to be $\alpha_{zz}=21.3$~\AA$^3$,
$\alpha_{yy}=14.5$~\AA$^3$, and $\alpha_{xx}=8.5$~\AA$^3$, where the
$z$-axis is parallel to the C-I axis, the $y$-axis is perpendicular to
the $z$-axis but still in the molecular plane, and the $x$-axis is
perpendicular to the molecular plane. It is therefore expected that a
linearly polarized laser field will align the C-I axis along the
polarization axis. With an elliptically polarized beam the largest
polarizability axis of the molecule is expected to be aligned along
the major axis of the elliptical field and the second most polarizable
axis along the minor axis of the field. Therefore, it is expected that
an elliptically polarized field will align the C-I axis along the
major polarization axis and the molecular plane to the polarization
plane. The case of such 1D and 3D alignment is discussed in
\autoref{alignment} and the alignment enhancement employing
state-selected molecules is demonstrated. The permanent electric
dipole moment of DFIB is 2.25~D. It is parallel to the C-I axis with
the negative end pointing in the direction of the iodine atom.
Consequently, it is expected that a static electric field, collinear
with the major polarization axis of the alignment field, will force
the iodine end of each molecule to point in the direction of the
positive electrode of the static field. The case of 1D and 3D
orientation and the advantage of using state selected molecules is
discussed in \autoref{orientation}.

\section{Experimental setup}
\label{exp-setup}

\begin{figure}
   \includegraphics[width=\figwidth]{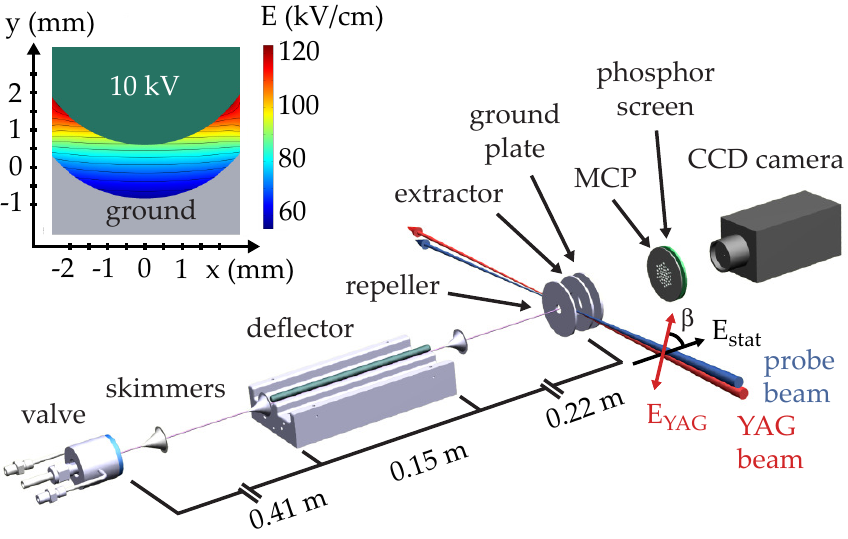}
   \caption{(Color online) Scheme of the experimental setup. In the
      inset, a cut through the deflector is shown and a contour-plot
      of the electric field strength is given. $\beta=0$ corresponds
      to the case of the major axis of laser polarization oriented
      along the TOF axis.}
   \label{fig:1}
\end{figure}

A schematic of the experimental setup, which is described in detail
elsewhere~\cite{Filsinger:deflection:inprep}, is shown in
\autoref{fig:1}.
A mixture of a few mbar of DFIB and 90 bar of helium is expanded into
vacuum using a pulsed Even-Lavie valve~\cite{even:2003:jcp} to produce
a molecular beam with a rotational temperature of approximately 1~K.
After passing two 1-mm-diameter skimmers the molecular beam enters a
15-cm-long electrostatic beam deflector. The deflector consists of a
trough (at ground potential) with an inner radius of curvature of
3.2~mm and a rod (at high voltage) with a radius of 3.0~mm. The
vertical separation of the two electrodes across the molecular beam
axis is 1.4 mm. This electrode geometry creates a two-wire field with
a nearly constant gradient over a large area around the molecular beam
axis~\cite{Ramsey:MolecularBeams:1956}. In the inset of
\autoref{fig:1}, a cross sectional view of the deflector with the
created electric field is given. The deflector is mounted such that
the deflection occurs vertically, and molecules in high-field-seeking
states are deflected upwards.

After passing the deflector, the molecular beam enters the
target/detection area through a 1.5-mm-diameter skimmer where it is
crossed by one or two laser beams that are focused by a spherical lens
with a focal length of $f=300$~mm. The lens is mounted on a vertical
translation stage so that the height of the laser foci can be adjusted
with high precision. One laser beam, consisting of 30-fs-long pulses
(800 nm, beam-waist of $\omega_{0}=21$~$\mu$m which corresponds to
$4\times10^{14}$~W/cm$^{2}$) is used to probe the molecules. In the
first part of the experiment, this laser beam is used to characterize
the deflection by determining the density, at a given height, of the
molecular beam via photoionization. Hereafter, this laser beam is used
for Coulomb exploding the molecules to determine their alignment and
orientation by imaging the ionized fragments on the plane of the
microchannel plate (MCP) detector. The second laser beam, consisting
of 10-ns-long pulses from a Nd:YAG laser ($1064$~nm,
$\omega_{0}=36$~$\mu$m), is used to align and orient the molecules.
Two intensity values of the pulses were applied: $1.2\times10^{12}$
and $1.8\times10^{11}$~W/cm$^{2}$, denoted hereafter as
I$_\text{YAG,high}$ and I$_\text{YAG,low}$, respectively.

The probe pulse is electronically synchronized to the peak of the YAG
pulse. Positive ions produced by the probe pulses are accelerated, in
a velocity focusing geometry, towards a MCP detector backed by a
phosphor screen. The 2D ion images are recorded with a CCD camera. In
particular, the I$^{+}$ and F$^{+}$ fragment ion distributions are
useful experimental observables since these ions recoil along the
symmetry axis and in the plane of the molecule, respectively. Thereby,
they allow the determination of the molecular orientation at the time
of the probe pulse. Also, the time-of-flight (TOF) of the ions from
the laser-interaction spot to the MCP is recorded. The experiments are
conducted at 20 Hz, limited by the repetition rate of the YAG laser.

\section{Results and Discussions}
\label{results}

\subsection{Deflection of the molecular beam}
\label{deflection}

\begin{figure}
   \centering
   \includegraphics[width=\figwidth]{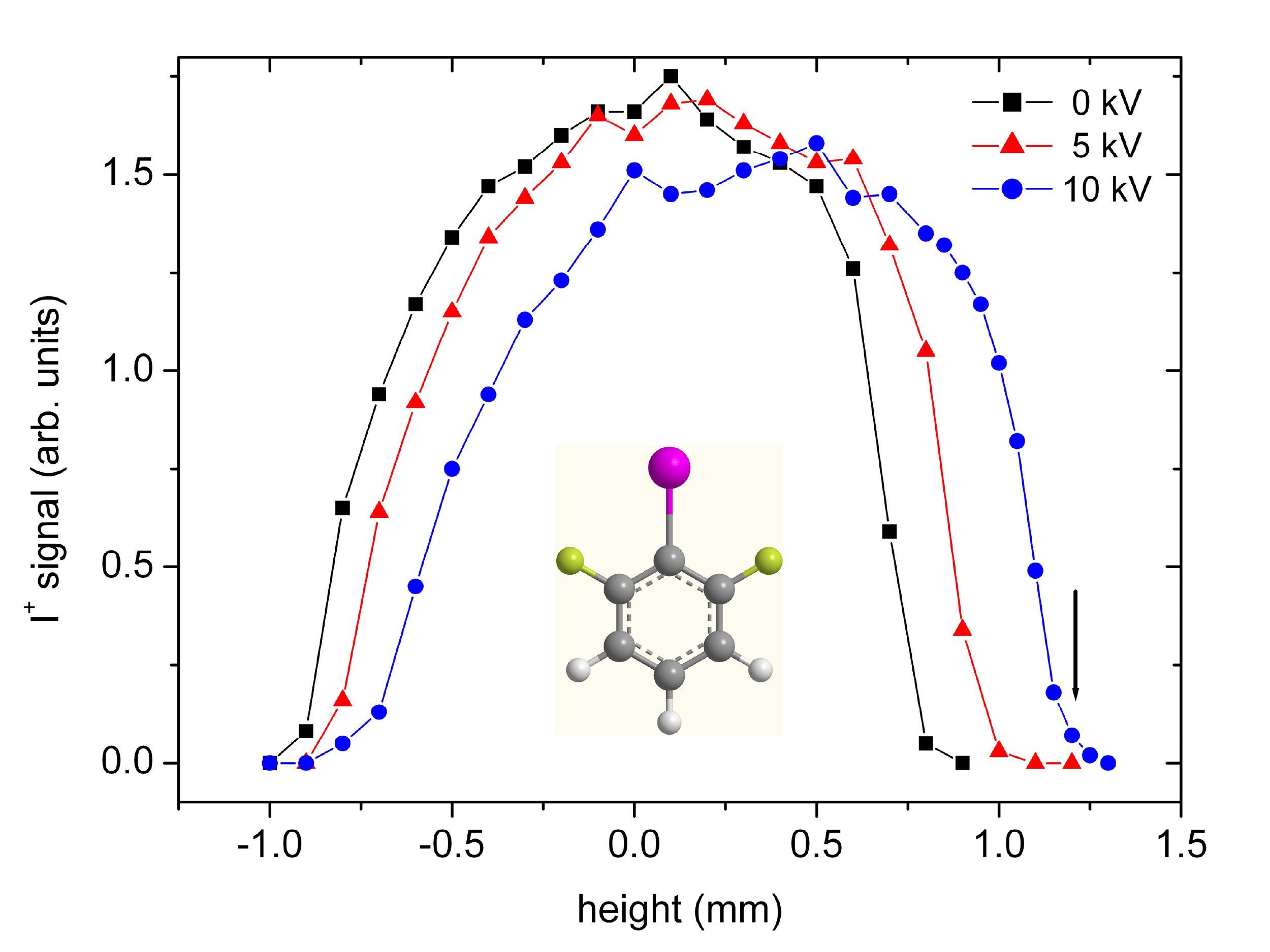}
   \caption{(Color online) Vertical profiles of the molecular beam
      measured by recording the I$^{+}$ signal as a function of the
      vertical position of the fs probe beam focus. The experimental
      data are shown by black squares (deflector off, 0~kV), red
      triangles (5~kV), and blue circles (10~kV). The arrow denotes
      the position of the probe laser focus for acquiring images and
      TOF spectra of deflected molecules. A model of the DFIB molecule
      is shown as an inset.}
   \label{fig:2}
\end{figure}

The effect of the deflector on the molecular beam is illustrated in
\autoref{fig:2} by the vertical intensity profiles, which are obtained
by recording the I$^{+}$ signal from photoionization due to the
femtosecond probe laser as a function of the vertical position of the
laser focus.
When the deflector is turned off, the molecular beam extends over
$\sim1.7$~mm, mainly determined by the diameter of the last skimmer
before the target area. When the deflector is turned on, the molecular
beam profile shifts upwards (by 0.4~mm at 10~kV). The shift becomes
more pronounced as the voltage on the deflector is increased from 5~kV
to 10~kV. Molecules in the lowest rotational quantum states have the
largest Stark shifts and are, therefore, deflected the most, as shown
in our recent work on iodobenzene
molecules~\cite{Holmegaard:PRL102:023001,
   Filsinger:deflection:inprep}. In the measurements described below,
experiments have been conducted on these quantum-state selected
molecules simply by positioning the laser foci close to the upper
cut-off region in the 10~kV profile, as indicated by the arrow in
\autoref{fig:2} ($height~=~1.2~mm$).

\subsection{1-dimensional and 3-dimensional alignment}
\label{alignment}

\begin{figure}
   \centering
   \includegraphics[width=\figwidth]{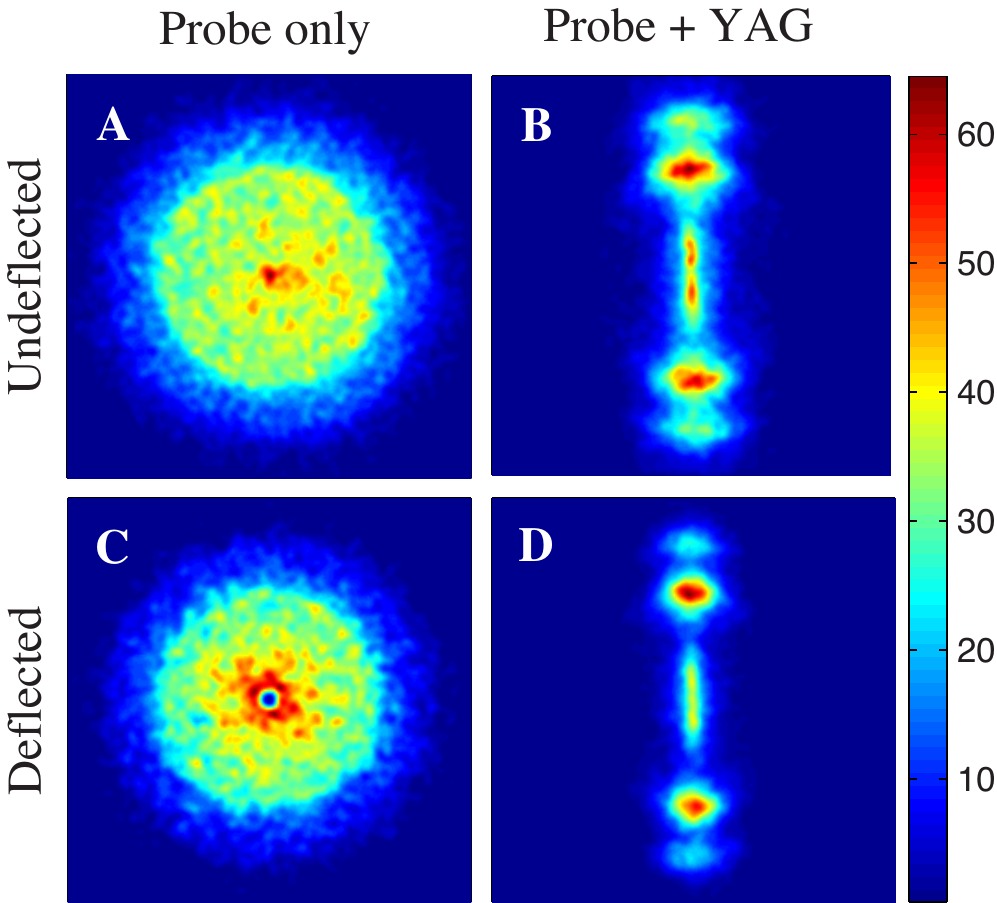}
   \caption{(Color online) Alignment of DFIB molecules induced by a
      linearly polarized YAG pulse, illustrated by velocity map images
      of the I$^+$ ions. A, C: I$^+$ images recorded without the YAG
      pulse for undeflected and deflected, respectively. B, D: I$^+$
      images with the YAG included (vertical polarization,
      I$_\text{YAG,high}$) using undeflected and deflected molecules,
      respectively.}
   \label{fig:4}
\end{figure}

We start by showing that a linearly polarized YAG pulse induces 1D
alignment of the DFIB molecules. We compare I$^{+}$ images recorded
with and without the YAG pulse when the deflector is turned off. In
the absence of the YAG pulse, the I$^{+}$ image is circularly
symmetric (\autoref{fig:4}~A) due to the circular symmetry introduced
by polarizing the probe pulse perpendicular to the detector plane.
When the YAG pulse, polarized parallel to the detector plane, is
applied, the I$^{+}$ ions exhibit strong angular localization about
the polarization axis, as depicted in \autoref{fig:4} B, showing that
the C-I axis of the molecules is strongly confined along the YAG
polarization, \ie, the DFIB molecules are 1D adiabatically aligned.
Consequently, images taken with the YAG polarization in the detector
plane can be regarded as a ``side-view'' of the molecule. The degree
of alignment is quantified by
$\langle\cos^2\theta_\text{2D}\rangle=0.94$, where $\theta_\text{2D}$
is the angle between the projection of the I$^+$ recoil velocity on
the detector plane and the YAG
polarization~\cite{kumarappan:2006:jcp}. The I$^{+}$ ions appear as
two sets of angularly narrow rings. The innermost (and brightest) ring
results from I$^+$ ions when DFIB is doubly ionized by the probe pulse
and fragments into an I$^+$~+~C$_6$H$_3$F$_2$$^+$ ion pair, whereas
the outermost ring results from I$^+$ ions formed from triple
ionization and fragmentation into an I$^+$~+~C$_6$H$_3$F$_2$$^{2+}$
ion pair~\cite{larsen:1999:jcp}. The value of
$\langle\cos^2\theta_\text{2D}\rangle$ is determined in the radial
range corresponding to the outermost ring. By comparison, in the
absence of the YAG pulse $\langle\cos^2\theta_\text{2D}\rangle=0.50$,
as expected for randomly oriented molecules. When the deflector is
turned on and the laser foci moved close to the upper cutoff region in
the 10 kV profile ($height~=~1.2$~mm), corresponding to the lowest
lying rotational states of the DFIB molecules, the 1D alignment
sharpens. This is evident by comparing \autoref{fig:4}~B and D and
noting that $\langle\cos^2\theta_\text{2D}\rangle$ increases to
$0.96$.
The improved alignment is consistent with our recent experiments on
iodobenzene~\cite{Holmegaard:PRL102:023001,Filsinger:deflection:inprep}.

\begin{figure}
   \centering
   \includegraphics[width=\figwidth]{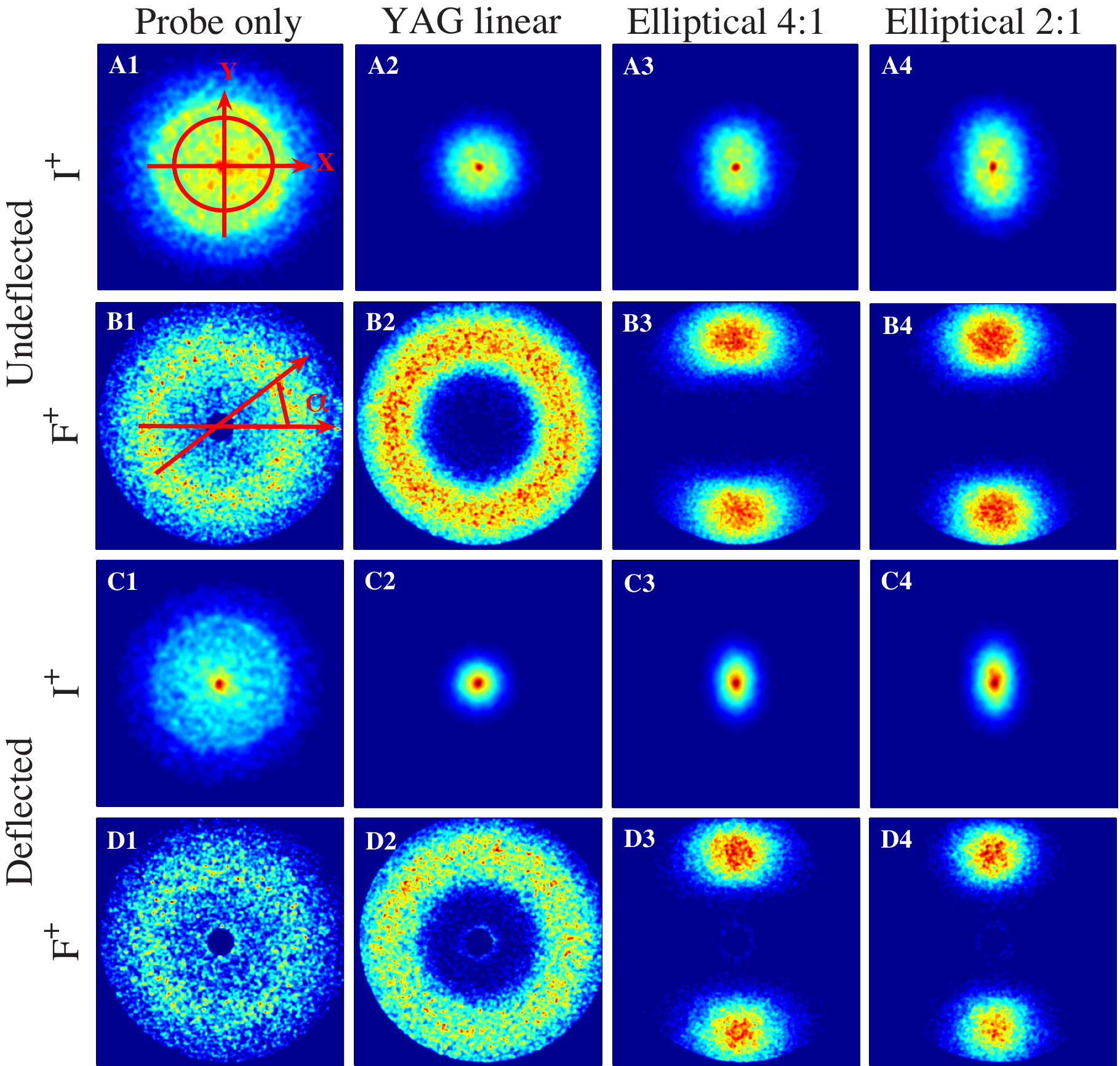}
   \caption{(Color online) Ion images of I$^{+}$ and F$^{+}$ fragments
      recorded in end-view and at I$_\text{YAG,high}$ demonstrating 1D
      and 3D alignment. Columns: (1) probe only; (2) YAG linearly
      polarized; (3), (4) YAG elliptically polarized with an
      ellipticity intensity-ratio of 4:1 and 2:1, respectively. Rows:
      (A) I$^{+}$ images from undeflected molecules (0~kV); (B) F$^+$
      images from undeflected molecules; (C) I$^+$ images from
      deflected molecules (10~kV); (D) F$^+$ images from deflected
      molecules.}
   \label{fig:5}
\end{figure}

Similarly, images taken with the YAG pulse polarized perpendicular to
the detector plane show ``end-views'' of the molecules. All images in
\autoref{fig:5} are recorded in the end-view.
The tight confinement of the I$^{+}$ ions near the center (A2--A4 and
C2--C4), compared to the I$^{+}$ distribution without the YAG pulse
(\autoref{fig:5}~A1 and C1) is an alternative way to visualize the 1D
alignment. In order to quantify the angular confinement, the average
distances $\langle{X}\rangle$ and $\langle{Y}\rangle$ of I${}^+$ ions
along the horizontal ($X$) and vertical axes ($Y$) from the center of
the end-view images are calculated, see \autoref{fig:5}~(A1). Without
the YAG $\langle{X}\rangle=37.4$~pixels and
$\langle{Y}\rangle=34.8$~pixels. Including the YAG
$\langle{X}\rangle=19.0$~pixels and $\langle{Y}\rangle=19.0$~pixels
without voltages applied to the deflector (Image A2 in
\autoref{fig:5}), and $\langle{X}\rangle=12.3$~pixels and
$\langle{Y}\rangle=12.2$~pixels with 10~kV applied to the deflector
(\autoref{fig:5}~C2).

The 1D alignment is also visible in the F$^{+}$ images. In the absence
of the YAG pulse the F$^{+}$ image, recorded in the end-view, takes
the form of a circularly symmetric ring (\autoref{fig:5}~B1 and D1).
This is caused by the fact that the linearly polarized probe field
preferentially ionizes molecules with their C-I axis along its
polarization vector \cite{larsen:1999:jcp}. When the linearly
polarized YAG pulse is applied, the circular symmetry is conserved and
the ring structure becomes more sharply defined to a localized radial
area with no signal detected in the central region (\autoref{fig:5}~B2
and D2). Such a ring structure is only compatible with the C-I axis
being sharply aligned perpendicular to the detector plane and with the
rotation of the molecular plane uniformly distributed around the C-I
axis.

\begin{figure}
   \centering
   \includegraphics[width=\figwidth]{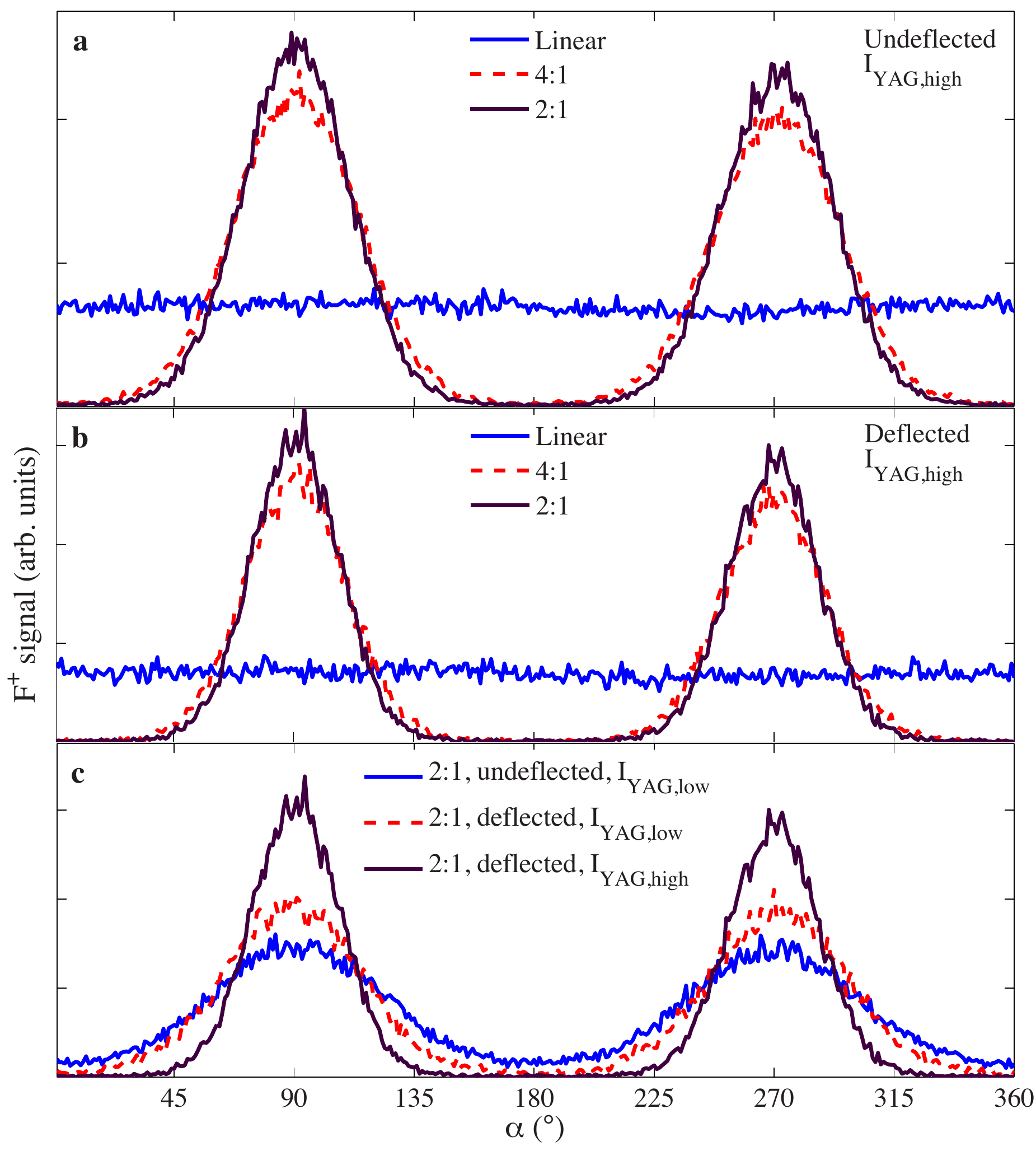}
   \caption{(Color online) Angular distributions of the F$^+$
      fragments for different YAG polarizations and intensities
      ($\alpha$ is pictorially defined in \autoref{fig:5}~B1). (a)
      undeflected molecules under I$_\text{YAG,high}$ (b) deflected
      molecules under I$_\text{YAG,high}$; (c) comparison between the
      angular distributions with I$_\text{YAG,low}$ and
      I$_\text{YAG,high}$ for deflected molecules, where all the
      curves correspond to a 2:1 ellipticity ratio.}
   \label{fig:6}
\end{figure}

Next, we investigate the effect on the molecular alignment when the
YAG polarization is changed from linear to elliptical. The end-view of
the F$^{+}$ ions shows a dramatic change [\autoref{fig:5}~B3, B4, D3,
D4]. The initial circular symmetry is replaced by sharp localization
along the minor axis (vertical in \autoref{fig:5}) of the elliptically
polarized YAG pulse. The angular localization of the F$^{+}$ ions,
largest for the 2:1 ellipticity ratio (B4 and D4), shows that the
molecular plane is confined to the plane defined by the elliptical
polarization, and the conservation of their radial confinement shows
that the C-I axis remains localized around the major polarization
axis. To quantify the planar confinement the angular distributions of
the F$^{+}$ ions were determined by radial integration of the images.
The results are displayed in \autoref{fig:6}.
Focusing on panel (a), containing the undeflected data, it is seen
that for the linearly polarized YAG pulse a uniform distribution
emerges, whereas for elliptical polarization the angular distribution
localizes. For the 4:1 ellipticity ratio (meaning that the intensity
along the major axis is four times the intensity along the minor axis)
the full widths at half maximum (FWHM) of the peaks in the angular
distribution are 52.5\degree; for the 2:1 ratio they are 48.5\degree.
Regarding the localization of the C-I axis, more precise insight is
obtained by observing the I$^{+}$ end-view images. It is seen that the
initial circular distribution (\autoref{fig:5}~A2) develops into an
elliptical shape when the YAG polarization is changed from linear to
elliptical. This shows that the C-I axis has suffered a small
distortion along the minor axis, remaining, however, tightly confined.
The distortion is largest for the 2:1 ellipticity ratio. This is
evident directly from the images as well as from the $\langle X
\rangle$ and $\langle Y \rangle$ values. With the 4:1 ratio (A3)
$\langle X \rangle$ = 18.1~pixels and $\langle Y \rangle$ =
21.0~pixels, whereas for the 2:1 ratio (A4) $\langle X \rangle$ =
18.2~pixels and $\langle Y \rangle$ = 23.4~pixels. Returning to the
F$^{+}$ images the C-I axis distortion is actually visible by slightly
more radial smearing in the 2:1 image compared to the 4:1 image. Thus,
we conclude that the elliptically polarized YAG pulse induces
pronounced 3D alignment of the DFIB. The confinement of the molecular
plane increases as the polarization state is brought closer to
circular polarization, but also gives rise to a larger distortion of
the linear confinement of the C-I axis. These findings are consistent
with previous observations on other molecules~\cite{larsen:2000:prl,
   simon_thesis}.

When the deflector is turned on the 3D alignment improves. The F$^{+}$
ion images show stronger confinement for both the 2:1 and the 4:1
configuration (\autoref{fig:5}~D3, D4) compared to the corresponding
undeflected data (\autoref{fig:5}~B3, B4). This is also clear from the
angular distributions in panel (b) of \autoref{fig:6}. For the 4:1
ratio the FWHM decreases from 52.5\degree\ to 43.0\degree and for the
2:1 ratio from 48.5\degree\ to 37.0\degree. Turning to the I$^+$
end-view images in row C it is seen that although the C-I axis is
still distorted when an elliptically YAG pulse is employed, the C-I
axis confinement is much improved compared to the situation for
undeflected molecules - compare \autoref{fig:5} C3, C4 to A3, A4. With
the deflected molecules $\langle{X}\rangle=10.7$~pixels and
$\langle{Y}\rangle=15.3$~pixels for the 4:1 ratio and
$\langle{X}\rangle=10.9$~pixels and $\langle{Y}\rangle=17.3$~pixels
for the 2:1 ratio. We conclude that both, 1D and 3D alignment, is
significantly enhanced when using quantum-state selected molecules.

\subsection{1-dimensional and 3-dimensional orientation}
\label{orientation}

Our final task is to prove that the molecules are oriented in addition
to being aligned. This is accomplished by supplementing the ion images
already presented by time of flight measurements of the I$^+$ ions.
The most optimal information is obtained by using the end-view
geometry. In this situation, the C-I axis is aligned parallel to the
static electric field from the electrodes (693 V/cm) in the ion
spectrometer. The TOF spectra of the I$^+$ ions will now allow us to
conclude that the permanent dipole moment (directed from the carbon
atom on the $C_2$ axis to the I atom) is preferentially pointing
towards the repeller plate (high electrostatic potential) as we
expect.

\begin{figure}
   \centering
   \includegraphics[width=\figwidth]{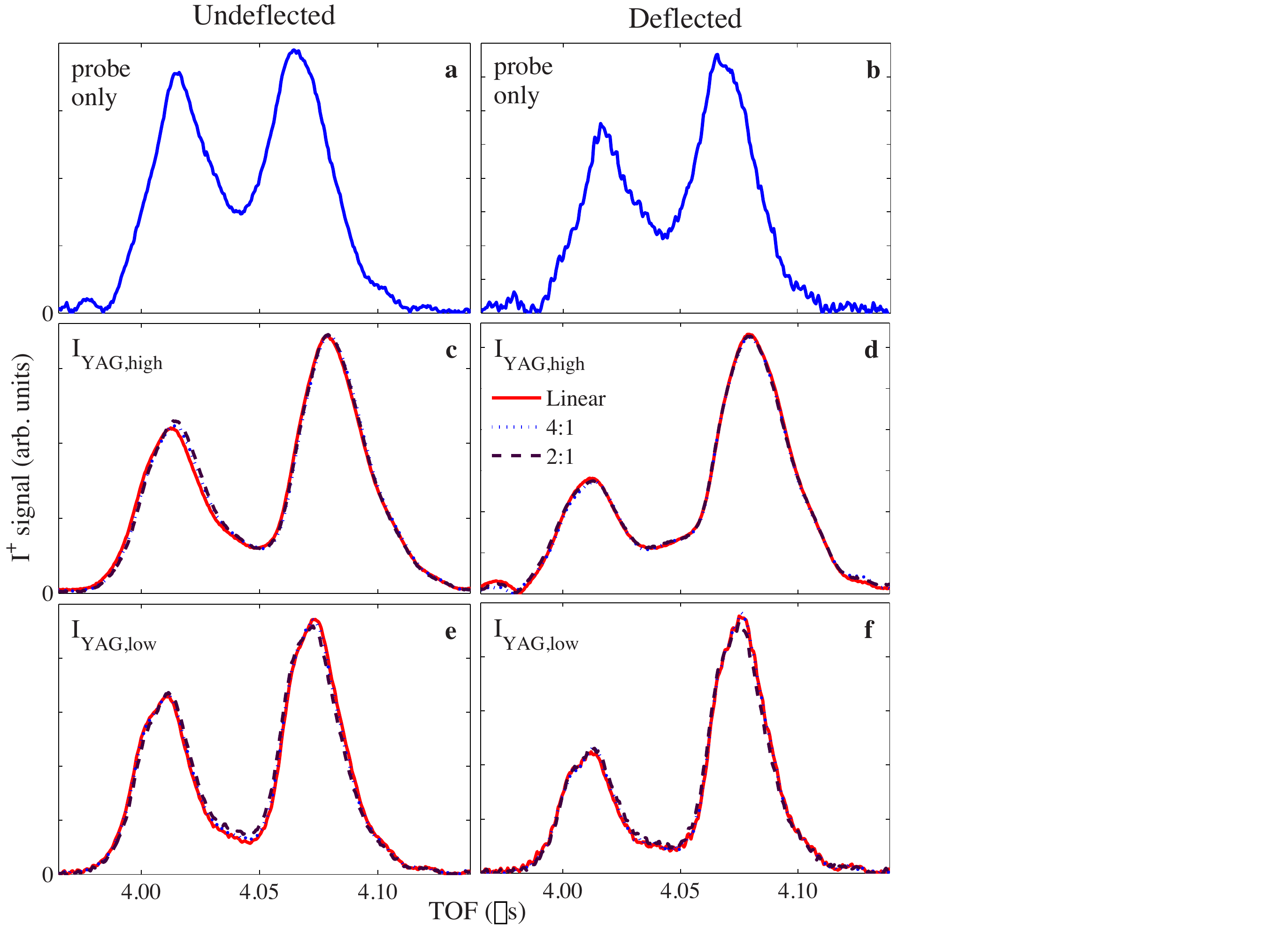}
   \caption{(Color online) Time-of-flight spectra of the I$^{+}$ ions
      recorded under conditions of (a)--(b): probe pulse only;
      (c)--(d): high intensity YAG pulse (I$_\text{YAG,high}$);
      (e)--(f), (c)--(d): low intensity YAG pulse
      (I$_\text{YAG,high}$); In panel (c)--(f) the YAG polarization is
      linear (solid red curves), 4:1 elliptical (dotted blue curves),
      and 2:1 elliptical (dashed black curves). The data in the left
      column are recorded for undeflected and the data in the right
      column for deflected molecules.}
   \label{fig:7}
\end{figure}

The TOF spectra recorded with the probe pulse alone are shown in
\autoref{fig:7}~(a) and (b).
As mentioned in connection with the F$^+$ ion images the linearly
polarized pulse is preferentially ionizing molecules with their axis
along its polarization axis which explains the appearance of a forward
and a backward peak~\cite{dietrich:1993} centered around at
4.045~$\mu$s. A mild asymmetry in the signal strength of the forward
and the backward peak is observed. For undeflected molecules the
asymmetry, defined as I$^+_b$/I$^+_f$ is $52/48$. For the deflected
molecules the asymmetry increases to $59/41$.

To get an insight into the origin of this asymmetry we calculate the
molecular orientation due to the static electric field itself. First,
the ensemble-averaged orientation cosine $\langle\cos\theta\rangle$ is
calculated quantum mechanically, where $\theta$ is the polar angle
between the molecular C-I axis and the static field direction. For an
ensemble at a rotational temperature of 1~K
$\langle\cos\theta\rangle=0.011$, whereas for an ensemble
corresponding to the state-selected molecules after the deflector
($height=1.2$~mm, see \autoref{fig:2})
$\langle\cos\theta\rangle=0.079$~\footnote{Since we have not performed
   a full simulation of the deflection profiles, as is described in
   reference~\onlinecite{Filsinger:deflection:inprep}, we have used
   the populations at the appropriate relative intensity value from
   the simulations on iodobenzene presented there.}. To obtain the
backward-forward ratio, $b/f$, we approximate the orientational
distribution function~\cite{fh:1991:jpc} by the truncated series
$n(\theta)=1+a_1\cdot{}P_1(\cos\theta)=1+3\cdot\langle\cos\theta\rangle\cdot\cos\theta$.
Higher order Legendre polynomials can be neglected for the low
electric field strength employed here. Furthermore, we assume that the
molecules are probed with a detection efficiency scaling as
$\cos^2\theta$ to account for the alignment selectivity of the probe
pulse. Using the calculated values for $\langle\cos\theta\rangle$ we
get $b/f=52/48$ for the 1~K ensemble and $b/f=60/40$ for the
state-selected ensemble. Both numbers agree well with the
observations. We conclude that the observed backward forward asymmetry
in the time of flight spectra observed in the absence of the YAG pulse
is due to weak orientation induced by the static field, as well as due
to alignment selectivity of the probe beam.

When the YAG pulse is included two changes of the TOF spectra occur.
First the forward and the backward peaks become more separated. This
results from the strong alignment induced by the YAG pulse. All
molecules are confined near the TOF axis, which will cause I$^+$ ions
to be ejected directly towards or 180 degrees away from the detector,
thus causing the maximum separation between the forward peaks. Second,
the asymmetry increases. For the undeflected data (\autoref{fig:7}~c)
the backward-forward ratio is 63/37 and for the deflected data
(\autoref{fig:7}~d) 73/27. The TOF spectra for the elliptically
polarized YAG (both 4:1 and 2:1 ratio) are essentially identical to
the ones for the linearly polarized YAG. Therefore, we conclude that
both strong 1D and strong 3D orientation are achieved. The degree of
alignment, as well as the degree of orientation, improves considerably
when the deflected molecules are employed.

\subsection{Alignment and orientation at low laser intensity}
\label{low-intensity}

There are many applications that would benefit strongly from being
conducted on samples of 3D oriented molecules, \eg,
photoelectron-angular-distribution measurements or diffractive
imaging. Some of these applications may, however, also be negatively
influenced, by the (moderately) intense field from the YAG pulse - in
the worst case a certain application may not be possible at all when
performed in the presence of the intense YAG field. The ideal
situation is to obtain the 3D orientation under field-free conditions.
This may be possible by, for instance, using timed perpendicularly
polarized short pulses~\cite{Lee:2006:prl} or by turning off the long
YAG pulse rapidly compared to the inherent molecular rotation times.
In the case of 1D alignment and orientation it has been shown that
such a conversion from adiabatic alignment/orientation to field-free
alignment/orientation is possible by femtosecond truncation of the YAG
pulse \cite{stolow:2003:prl,Sussman:pra:2006,Goban:PRL:2008}. One
consequence of the rapid turn-off method, or of other short pulse
nonadiabatic alignment or orientation methods
\cite{ghafur:natphys:2009}, is that the alignment/orientation becomes
transient, which could reduce the utility for several important
applications.

\begin{figure}
   \centering
   \includegraphics[width=\figwidth]{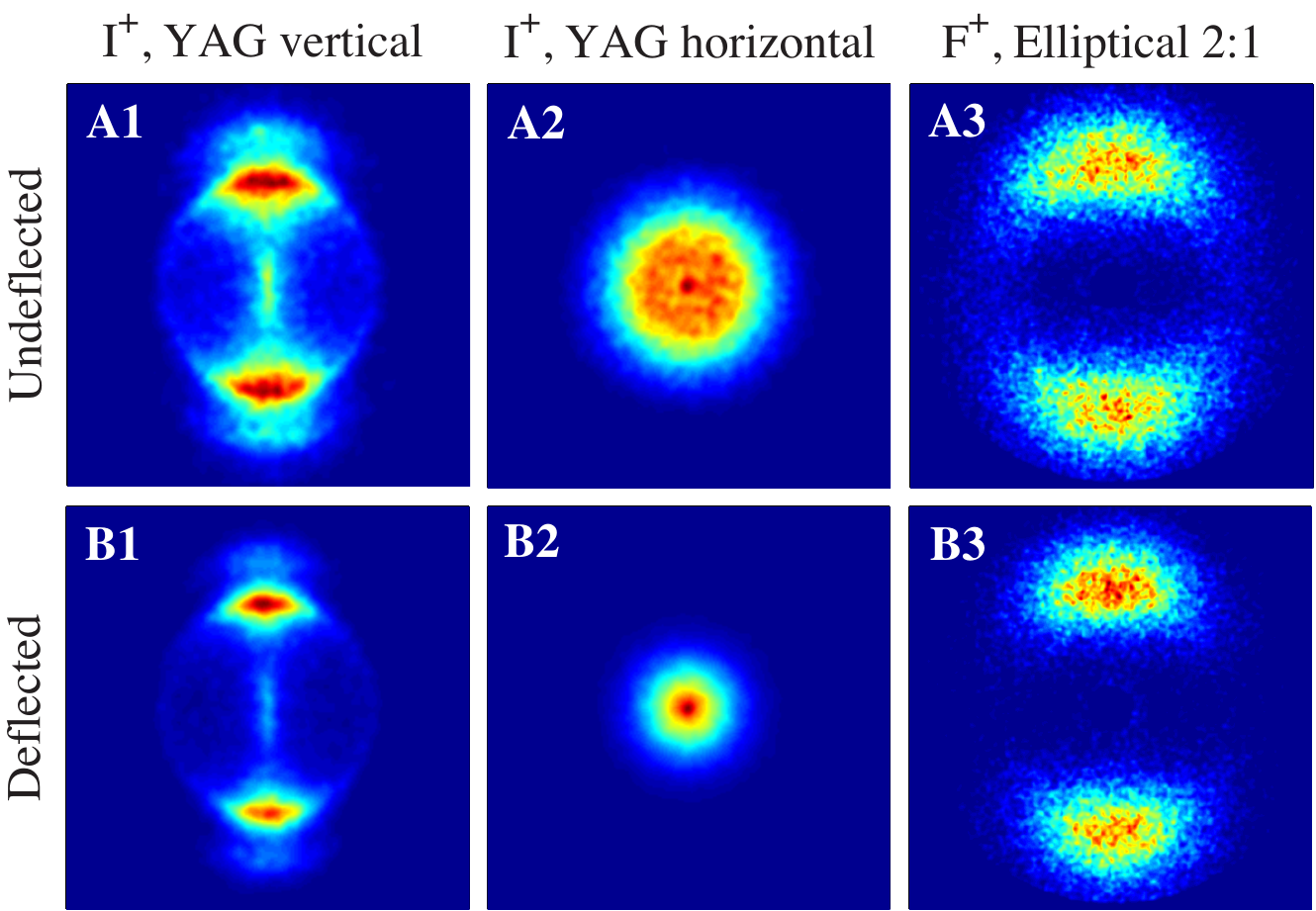}
   \caption{(Color online) Ion images of I$^+$ and F$^+$ fragments
      recorded under I$_\text{YAG,low}$. Columns: (1) I$^+$ side-view;
      (2) I$^+$ end-view; (3) F$^+$, ellipticity ratio of 2:1. Rows:
      (A) undeflected molecules; (B) deflected molecules.}
   \label{fig:8}
\end{figure}

In the present work, we decided to test if strong 3D alignment and
orientation could be maintained as the intensity of the YAG pulse was
lowered, since a weaker alignment field also would reduce its
influence in future applications. The ion images obtained at low YAG
intensities, shown in \autoref{fig:8}, reveal the same tendency as the
high intensity images shown in \autoref{fig:5}.
They show considerably stronger confinement of the C-I axis and of the
molecular plane, \ie, better 1D and 3D alignment, for deflected
molecules relative to the undeflected molecules. In detail, the
side-view image of the I$^{+}$ ions from the deflected molecules
(\autoref{fig:8}~B1) correspond to a value of
$\langle\cos^2\theta_\text{2D}\rangle=0.922$, and its end-view image
(\autoref{fig:8}~B2) shows a circular spot with
$\langle{X}\rangle=15.7$, and $\langle{Y}\rangle=15.8$, whereas for
the undeflected molecules $\langle\cos^2\theta_\text{2D}\rangle=0.874$
(\autoref{fig:8}~(A1)), and $\langle{X}\rangle=22.5$, and
$\langle{Y}\rangle=22.6$ (\autoref{fig:8}~A2).

The images of the F$^+$ ions (for a 2:1 ratio) exhibits stronger
angular confinement when deflected molecules (\autoref{fig:8}~C2)
rather than undeflected molecules (\autoref{fig:8} C1) are used. This
is also evident in the corresponding angular distributions in
\autoref{fig:6}~(c) that show a peak with a FWHM of 55.5\degree\ for
the deflected molecules relative to more than 72\degree\ for the
undeflected molecules. Finally, the TOF spectra, which indicate the
degree of orientation, show a backward-forward asymmetry of 58/42
(\autoref{fig:7}~e) for the undeflected molecules, which increases to
67/33 (\autoref{fig:7}~f) for the deflected molecules. In summary,
even when the intensity of the YAG alignment pulse is lowered by an
order of magnitude a strong degree of 3D alignment and orientation is
maintained, in accordance with previous findings for 1D alignment and
orientation~\cite{Holmegaard:PRL102:023001,
   Filsinger:deflection:inprep}.

\section{Conclusion}
\label{conclusion}

It was shown experimentally that 1-dimensional and 3-dimensional
alignment and orientation of an asymmetric top molecule is
significantly improved by selecting those molecules in a cold
molecular beam that reside in the lowest rotational states. Alignment
was induced in the adiabatic regime, where the turn-on and turn-off
time of the linearly (1D alignment) or elliptically (3D alignment)
polarized laser field occur slowly compared to the inherent rotational
periods of the molecules. In addition, orientation, induced by the
combined action of the laser field and a weak static electric field,
is taking place in the adiabatic regime.

The alignment and orientation were illustrated by studies on
2,6-difluoroiodobenzene. Here, the major polarizability axis is
parallel to the direction of the permanent dipole moment.
Consequently, 3D orientation was achieved by positioning the major
polarization axis of the elliptically polarized laser pulse along the
direction of the static electric field. The case of the permanent
dipole moment being parallel to the major polarizability axis occurs
for a large number of molecules, and we expect that the method will be
generally applicable for that case. Another possibility is that the
major polarizability axis is perpendicular to the dipole moment. We
believe that the elliptically polarized pulse should be sent such that
the minor polarization axis coincides with the direction of the static
electric field.

Finally, in the most general case, occurring for the largest range of
molecules, the major polarizability axis and the permanent dipole
moment differ by an angle that is neither zero nor ninety degrees.
Achieving 3D orientation for this class of molecules attracts special
interest since examples include important biomolecules such as amino
acids and nucleic acids. We believe that 3D orientation can be
achieved by turning the molecule, by rotating the axes of the
elliptically polarized laser pulse, such that the permanent dipole
moment coincides with the static field direction.

\begin{acknowledgments}
   We thank Mikael P.\ Johansson for calculating the polarizability
   tensor of DFIB. Iftach Nevo acknowledges support from the Marie
   Curie IntraEuoropean Fellowships Network FP6. We also acknowledge
   support from the Carlsberg Foundation, the Lundbeck foundation, the
   Danish Natural Science Research Council, and the Deutsche
   Forschungsgemeinschaft within the priority program 1116.
\end{acknowledgments}

\bibliography{ref}

\end{document}